\documentclass{icrc2009}

\usepackage{graphicx}   
\usepackage[caption=false]{caption}    
\usepackage[font=footnotesize]{subfig} 
\usepackage{fixltx2e}
\usepackage{url}

\newcommand{\shorttitle}[1]%
{\markboth{Proceedings of the 31\MakeLowercase{$^{st}$} ICRC, {\L}\'{o}d\'{z} 2009}{#1} }
\newcommand{\etal}{\MakeLowercase{\textit{et al. }}} 

\begin{document}
\title{Upper limits for pulsars with MAGIC (2005/2006 observations)}

\author{\IEEEauthorblockN{R. de los Reyes\IEEEauthorrefmark{1},
			  W. Bednarek\IEEEauthorrefmark{2},
                          M. Camara\IEEEauthorrefmark{1} and
                           M. Lopez\IEEEauthorrefmark{3} for the MAGIC collaboration}
                            \\
\IEEEauthorblockA{\IEEEauthorrefmark{1} Universidad Complutense, E-28040, Madrid, Spain}
\IEEEauthorblockA{\IEEEauthorrefmark{2} University of \L\'od\'z, PL-90236 Lodz, Poland}
\IEEEauthorblockA{\IEEEauthorrefmark{3} Universit\`a di Padova and INFN, I-35131 Padova, Italy}}

\shorttitle{R. de los Reyes  \etal Upper limits for pulsars with MAGIC}
\maketitle

\begin{abstract}
 During the last 3 years the MAGIC collaboration has searched for high-energy gamma-ray emission of some of the most promising pulsar candidates. 
The low energy threshold of MAGIC offered the opportunity for a high sensitivity search just above 50-100 GeV (since 2008 a new trigger system has decreased the energy threshold to 25 GeV), an energy region up to now inaccessible to ground based instruments and past satellite borne detectors. No pulsed gamma-ray emission has been detected from any pulsar observed during the observation campaigns of 2005-2006. 
Here we present  the upper limits obtained for two canonical pulsars (PSR J0205+6449 and PSR 
J2229+6114) and their host nebulae (3C58 and Boomerang, respectively) and the millisecond pulsar PSR J0218+4232. Physics implications will be discussed.
  \end{abstract}

 
\section{Introduction}
 
More than 2000 pulsars have been discovered at radio frequencies while at  
$\gamma$-ray energies EGRET detected only 7 high significance and 3 low significance emitters of pulsed $\gamma$-rays up to 5 GeV. On the other hand, no indication of pulsed emission in the energy band studied between 150 GeV and 50 TeV has been yet observed~\cite{Aharonian2007a, Albert2008c, Albert2007,Konopelko2007} by the latest generation of ground-based imaging air Cherenkov telescopes (IACT). Only recently, the energy range between 5-150 GeV starts to be covered by the satellite {\it Fermi} Gamma-Ray Observatory, AGILE and MAGIC IACT, which has detected pulsed $\gamma$-emission from Crab pulsar above E = 25 GeV within 22 h of  observation~\cite{CrabScience}. 
Moreover, the MILAGRO Collaboration reports 6.6$\sigma$ detection of the $\gamma$-ray source at energy $\sim 35$ TeV coincident with PSR J2229+6114~\cite{Abdo2009}, showing evidences of a possible extended emission. On the other site, {\it Fermi} Observatory has recently detected pulsed emission from $\sim30$ new $\gamma$-ray pulsars. Between them there are PSR J0205+6449 and PSR J2229+6114 (\cite{FermiCataloga}~\cite{Pellizzoni2009}).  
From the theoretical site, the constraints on the high energy emission from the pulsars and their nebulae above $\sim20$ GeV are very important in order to decide which proposed models are really at work. The popular models able to explain the pulsed emission from the inner pulsar magnetospheres 
(i.e. polar, outer and slop gap models) have been recently severely constrained  by the MAGIC detection of the pulsed emission from the Crab pulsar~\cite{Albert2007}. However, it is not clear at present whether the MAGIC discoveries in the case of the Crab pulsar are common for the whole 
$\gamma$-ray pulsar population. The models for the steady emission from the pulsar wind nebulae predict the $\gamma$-ray fluxes which are typically within the sensitivities of the current Cherenkov telescopes above $\sim 100$ GeV.  In the case of the PWNe around PSR J0205+6449 or PSR J2229+6114, the estimates based on the specific model for high energy radiation from PWNe~\cite{Bednarek2005}, are within the range of $\sim 1-4\%$ and $\sim 4-16\%$  of the Crab flux 
above $\sim 200$ GeV, respectively. However, surprisingly some of the best candidate PWNe expected as a TeV $\gamma$-ray emitters have not been detected up to now (e.g. nebulae around 
PSR B1951+32~\cite{Albert2007} or PSR 1706-44~\cite{Aharonian2007a}). On the other hand, 
it is quite interesting that some of the unidentified HESS TeV sources seem to be clearly related to the PWNe (e.g.~\cite{Aharonian2007b}).
We have selected the promising pulsar candidates for MAGIC observations by using a high spin-down power of the pulsar (E/d$^2$) as selection criterion instead of using the predictions of a specific model, since this selection criterion might also be valid for the best candidates of DC emission from the surrounding nebulae.
The best candidates were selected with the additional condition that the source culminates at not too large zenith angles, i.e. is well observable with a low threshold. 
Within the canonical pulsar population PSR J0205+6449 is, just after the Crab pulsar (PSR B0531+21), the most promising candidate followed by the EGRET detected pulsars Geminga (PSR J0633+1746), PSR B1951+32 and PSR J2229+6114. Within the millisecond pulsar population, PSR J0218+4232 is the first millisecond pulsar detected in $\gamma$-rays.
The MAGIC observations presented here were optimized for source searches above 60-100 GeV and were motivated to study the emission in the upper part of the GeV spectrum and possibly extending the observation into the TeV energy range. This observations can partially constrain outer gap models for pulsed emission and the steady state (DC) emission from the host nebulae at around 100 GeV.

 \section{Observations and data analysis}\label{sec:obs}
 \subsection{Observations}
MAGIC observations of the selected pulsars and the associated pulsar wind nebulae (PWN) have been carried out in 2005 and 2006 (up to January 2007). The 17-m single-dish MAGIC Cherenkov telescope~\cite{Lorenz2004} is located at the Roque de los Muchachos observatory site (28$^\circ$ 45Õ 34Ó N, 17 52Õ34ÓW, 2200 a.s.l.) on the Canary island of La Palma. An Active Mirror Control system yield a PSF of 0.025$^\circ$ for distant point-like light sources. 
The shower images recorded by a 3.2$^\circ$ FoV camera, consisting on 577 PMTs, are digitized by a 300 MHz Flash-ADC system in a time slice of 50 nsec, together with the absolute event time from a Rubidium clock synchronized via GPS giving a time accuracy of 200 nsec.  
MAGIC uses two different trigger concepts. The standard trigger, used for all observations until late 2007, is optimized for source searches above 60-100 GeV (zenith angle dependent) in a field of view (FOV) of about 1$^\circ$ radius around the candidate position, while a recently developed trigger logic~\cite{CrabScience} allows us to search for lower energy pulsed emission ($\ge$ 25 GeV) but in a more restricted FOV.
The observations presented here were carried out in the so-called ON-OFF mode during MAGIC cycle I (2005) and II (2006) by using the standard trigger~\cite{Albert2008c}. The data were analyzed, firstly, looking for summed $\gamma$-ray emission of the Pulsar and PWN generated in an area of the $\gamma$-ray PSF (DISP $<$ 0.10$^\circ$) of the telescope (ALPHA-analysis). Secondly, analyzing only the emission coming from the pulsar magnetosphere (Timing analysis). Finally we searched for excess events coming from the pulsar surroundings (0.1$^\circ$$<$ DISP $<$ 1$^\circ$).

\subsection{Data analysis}
The data are processed by the standard analysis program MARS~\cite{Bretz2005}, rejecting bad quality runs. This standard analysis includes a NSB background filter through a core-boundary image cleaning levels of 10-5 photoelectrons and $\gamma$/hadron separation by Random Forest (RF)~\cite{Albert2008c}. The cuts have been optimized by training on Crab data taken under the same conditions as the ON data to get the best telescope sensitivity for each energy bin. The energy resolution of reconstructed events, through the RF method, is 20-30 \% depending on the observationÕs zenith angle and incident energy. In the next analysis step one correlates the shower images with the source position by applying a cut in the Hillas parameter ALPHA~\cite{Hillas1985}. The OFF background distribution is normalized to the ON distribution for ALPHA values $>$ 30¡. The number of excess events is N$_{exc}$=N$_{On}$-N$_{bg}$ in the ALPHA range 0-10$^\circ$. 
The analysis of the data recorded in 2005 and 2006 allowed us to achieve within 50 hours of observation a sensitivity for a 5$\sigma$ excess in the case of a minimum DC flux of about $\sim$2.3\% of that of the Crab nebula\footnote{After some recent changes of the readout using a new multiplexed 2 GHz Flash ADC system allowing for a refined timing analysis the sensitivity could be improved in 2007 to a 1.4\% Crab flux limit in 50 hours observation time.}. 
The significance of a certain signal on top of some background in a certain data bin (in ALPHA- or energy) is obtained through the Li \& Ma expression (formula 17 of\cite{LiYMa1983}). 
In the case of pulsed emission, the sensitivity can be further enhanced if one knows the pulsar period.
 The integral flux upper limits are calculated following the Rolke method~\cite{Rolke2005} for a 99\% (3$\sigma$) confidence level considering the objects under study as point-like sources due to the emission measured at high energies. 
However, there could be emission from the vicinity of the pulsar due to possible interaction between the pulsar wind nebula and the interstellar medium~\cite{Albert2008b} or from interactions between the pulsar wind nebula and the IM due to the binary movement~\cite{Albert2008b}.  To look for emission in a wider region around the pulsar, we reconstruct the possible origin in the sky plane through the so-called DISP-algorithm~\cite{Domingo2006} with a reconstruction error of 0.1$^\circ$. The systematic pointing uncertainty due to unknown telescope deformation and tracking errors is estimated to be below 2'~\cite{Albert2008c}.

\subsection{Timing analysis}\label{sec:time}
To search for the pulsed emission from the magnetosphere of the pulsar we have performed a periodicity analysis of the data converting the events arrival time measured by the observatory through a GPS-Rubidium clock system\footnote{Tested with optical data of the Crab pulsar recorded with the MAGIC central pixel~\cite{Lucarelli2008}} to the pulsar reference frame (the solar system barycentre (SSB)~\cite{Taylor1989}).
These corrected events are folded to the pulsar phase (light curve) given by the pulsar ephemeris in X-rays (RXTE satellite) and radio observations (Nancay radio telescope).
The light curve of the pulsars studied here is characterized by a double-peaked structure; therefore we have used a Pearson $\chi^2$-test and a H-test~\cite{deJager1989} as statistical periodicity tests. 

Guided by the two main models of pulsar magnetosphere emission, the timing analysis has been carried out in two energy windows: the first one focused on low energies (E$<$300 GeV), while the second one includes all the events collected by the telescope in order to include the possible emission extended up to TeV energies. In both energy windows we have optimized the cut in the hadronness parameter for the $\gamma$/hadron separation by requiring that 80\% of the $\gamma$-rays in Monte Carlo simulations survive the cuts.

\section{Results}\label{sec:res}

\subsection{PSR J0205+6449 and the PWN 3C 58}

\begin{figure}[!t]
\centering
\includegraphics[width=3.0in]{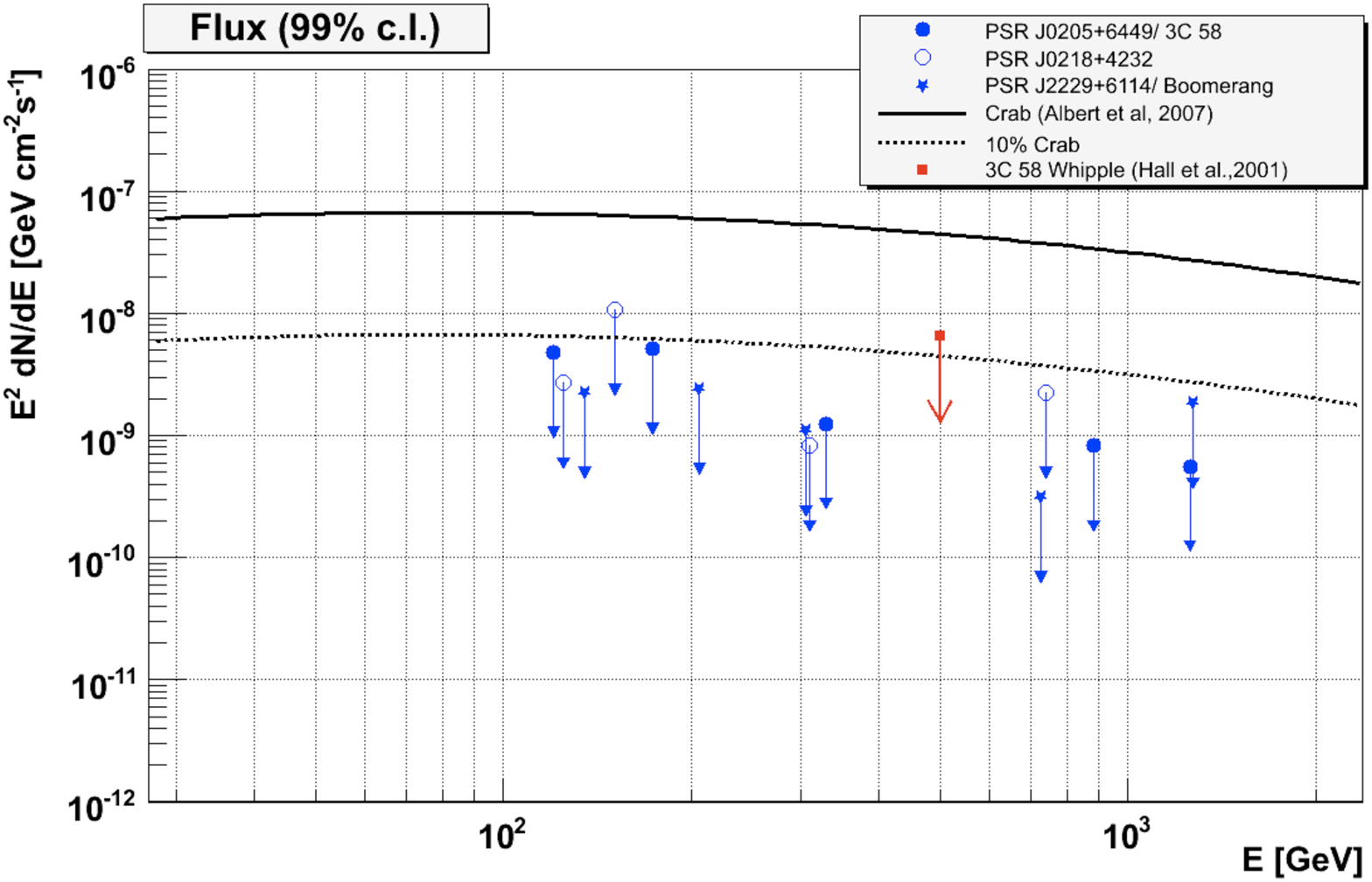}
\caption{\it DC Flux upper limits (99\% c.l.) for the three pulsars and their corresponding nebulae from MAGIC observations (blue arrows): 3C58 and PSR J0205+6449 (filled circles); Boomerang and PSR J2229+6114 (filled stars); and PSR J0218+4232 (empty circles). The black lines denote different fractions of the Crab flux: dashed line shows 10\% and solid line 100\% of the Crab flux . The red filled square is the upper limit obtained from Whipple observations of 3C 58~\cite{Hall2001}.}
\label{fig:flux}
\end{figure}

This pulsar has the second highest spin-down power of all members of the radio pulsar population, $\sim$ 0.02 times the one of the Crab pulsar, and a magnetic field strength at the neutron star surface of the same order of magnitude as Crab with a distance of 3.2 kpc. 
3C 58 was identified as a pulsar wind nebula (PWN) twenty years before its pulsar, PSR J0205+6449, which initially was discovered in X-rays~\cite{Murray2002} and only detected afterwards in radio frequencies~\cite{Camilo2002}. Observations of the PSR J0205+6449/3C 58 complex were carried out by MAGIC between 2005 September-December (MJD = 53625-53707) at zenith angles ranging between 36$^\circ$ - 45$^\circ$. 
After rejecting all data taken during bad weather conditions or non-perfect detector response, we collected 30 hours of PSR J0205+6449/3C58 data suitable for analysis. The flux limit for a threshold of 320 GeV to detect a 5$\sigma$ DC $\gamma$-ray signal corresponds to 2.8\% of the Crab flux. No significant excess signal (1.0$\sigma$) has been observed for the total energy range measured down to 100 GeV. The upper limits for different energy bins (in GeV.cm$^{-2}$.s$^{-1}$) are shown in figure~\ref{fig:flux} including the Whipple flux upper limit for E $>$ 500 GeV~\cite{Hall2001}. The integral flux upper limit (99\% c.l.) of PSR J0205+6449/3C 58 for the complete energy range above 100 GeV is listed in table~\ref{tab:steady}.
The extension of the emission region of 3C 58 is up to 6' in radio frequencies~\cite{Bietenholz2006}, which still corresponds to be point-like in case of the MAGIC angular resolution. A wider search in the pulsar surroundings has yielded no evidence of emission coming the interaction of the PWN with the interstellar medium at different energy integral bins. 
An insignificant excess of  Å 0.2$\sigma$ in the phase diagram at the expected peak phase positions has been obtained for the contemporaneous ephemeris from PSR J0205+6449 to the MAGIC observations taken by the monitoring of the RXTE satellite~\cite{Ransom2004}.
The upper limit to the pulsed emission of PSR J0205+6449 at this confidence level is F$_{ul}^{3\sigma}$  (E$_{th}$ = 280 GeV) $<$ 6.5 10$^{-13}$ cm$^{-2}$.s$^{-1}$, which includes $\gamma$-rays above 110 GeV.


\subsection{PSR J2229+6114 and the Boomerang PWN}

This pulsar was discovered in 2001 both in radio (Jodrell Bank Observatory) and in X-Rays (Chandra satellite)~\cite{Halpern2001}, inside the error box of the EGRET source 3EG J2227+6122, and within PWN G106.6+2.9. It has the 10th highest spin-down power of the MAGIC candidates considering its distance $\sim$3 kpc~\cite{Halpern2001}. MAGIC observed this pulsar and its nebula, so-called Boomerang, during 2005, August to December (MJD = 53584 - 53708) at zenith angles ranging between 32$^\circ$ and 38$^\circ$. For 10.5 hours of good data and a threshold of 196 GeV we expect a flux sensitivity of 3.4\% compared to the Crab flux. For the total range of accessible energies (E$>$100 GeV), the significance of the excess events is 0.6$\sigma$. The upper limits for selected energy bins (in GeV cm$^{-2}$ s$^{-1}$) are shown in figure~\ref{fig:flux} and its integral upper limit is listed in table~\ref{tab:steady}. A DISP-analysis has revealed no emission in the surroundings of this point-like source. 
No significant signal from PSR J2229+6114 has been found for any energy range using the most recent ephemeris~\cite{Halpern2001} to the MAGIC observations. The upper limit for a possible pulsed emission has been calculated using RolkeÕs method requiring a confidence level of 99\%. The integral value for the total energy range is F$_{ul}^{3\sigma}$  (E$_{th}$ = 300 GeV) $<$ 2.6 10$^{-12}$ cm$^{-2}$s$^{-1}$, which includes $\gamma$-rays above 100 GeV. 

  \begin{table}[!h]
\caption{Flux upper limits for the quoted energy bins E.}
  \label{tab:steady}
  \centering
  \begin{tabular}{l c c c c}
  \hline
 Pulsar  &  E$_{th}$ & N$_{exc}$ &      S      & F$_{3\sigma}$ (x10$^{-12}$)\\
 (PSR)  &   (GeV)  &           &          ($\sigma$) & (cm$^{-2}$s$^{-1}$) \\
 \hline
   \hline 
J0205+6449/3C 58  &  280  & 51 &  1.0 & 7.7 \\
J2229+6114/Boomerang  & 300  & 22 &  0.6 & 3.95 \\
J0218+4232 & 140  & 49 &  0.4 & 31.7 \\
  \hline
  \end{tabular}
  \end{table}

\subsection{PSR J0218+4232}

Discovered at radio wavelengths~\cite{Navarro1995}, the 2.3 ms pulsar PSR J0218+4232 belongs to a 2-day orbital period binary system with a low mass (0.2 M$_{\odot}$) white dwarf companion. PSR J0218+4232 is one of the three known millisecond pulsars (together with PSR B1821-24 and PSR B1937+21) with a hard spectrum and high luminosity emission in X-rays and the only one among them seen at  $\gamma$-ray energies up to energies of 1 GeV before the launch of the {\it Fermi} Observatory. Above 1 GeV the EGRET detection is consistent, within the error box, with the nearby AGN 3C 66A~\cite{Kuiper2000}. MAGIC observed PSR J0218+4232 at zenith angles between 14$^\circ$ and 32$^\circ$ during 2006 October and 2007, January (MJD = 54010 - 54115).  After excluding data recorded during adverse weather conditions and non-perfect detector performance the data from 20 hours observation time were further analyzed. The analysis sensitivity for PSR J0218+4232 above 200 GeV is, in units of the Crab flux, 3.3\%. 20 hours of analyzed data yielded no significant results (0.4$\sigma$). The corresponding upper limits (for 99\% CL) for different energy bins (in GeV.cm$^{-2}$.s$^{-1}$) are shown in figure~\ref{fig:flux} and the integral upper limit is listed in table~\ref{tab:steady}.
Although the pulsar phase diagram has a large DC emission, no pulsar wind nebula emission has been observed from the direction of the source outside 1" diameter from the pulsar position~\cite{Kuiper2002}. 
Therefore this source is considered to be point-like in the MAGIC analysis and no emission coming from possible interactions with the IM has been detected.
Although the millisecond pulsars are characterized to be very stable rotators, we got ephemeris from the Nancay radio telescope contemporaneous to the MAGIC observation period to improve the significance. From the timing analysis of $\gamma$-rays above 70 GeV coming from PSR J0218+4232, we obtain an upper limit of F$_{ul}^{3\sigma}$  (E$_{th}$ $>$140 GeV) $<$ 9.4 10$^{-12}$ cm$^{-2}$.s$^{-1}$.

\section{The consequences of PWNe upper limits}\label{sec:pwn}

The flux upper limits on the steady emission from the nebulae around two classical radio pulsars
reported in this paper are close to the estimates based on the model for the TeV $\gamma$-ray emission from PWNe discussed by Bednarek \& Bartosik~\cite{Bednarek2005}.
In the case of the Boomerang nebula around PSR  J2229+6114, the predicted flux
above 200 GeV is $\sim 4\%$ of the Crab Unit (CU) for the case of IC scattering of synchrotron and microwave background radiation. 
Bednarek \& Bartosik considered also the model with significantly stronger low energy background inside the nebula caused by the additional infrared component as expected in the case of the nebula around PSR 1706-44. 
In this case the estimated flux above 200 GeV is $\sim 16\%$ of CU. 
Our observations definitively exclude the presence of such additional radiation field inside the Boomerang nebula since the MAGIC upper limit at $\sim 200$ GeV is clearly lower. 
Considering the distance of PSR J2229+6114 of around 0.8 kpc~\cite{Kothes2001}, the flux estimated by Bednarek \& Bartosik should be an order of magnitude larger. In this case also the model with low soft radiation background inside the Boomerang nebula is in contradiction with the MAGIC upper limit.
The MAGIC upper limit at $\sim 1$ TeV of the pulsar PSR J2229+6114 and the recent detection of MILAGRO of an extended $\gamma$-ray source coincident with this source at $\sim 35$ TeV~\cite{Abdo2009} allows us to constrain the general spectral shape over above two decades above $\sim 100$ GeV. It looks that the spectrum should be relatively flat with the differential spectral index lower than $\sim2$, which is very different from that one observed in the case of the Crab Nebula.
This suggests that the acceleration and radiation processes inside Boomerang nebula can differ 
significantly from those ones observed in the Crab Nebula.
In the case of the nebula 3C 58 around PSR  J0205+6449, the flux estimated by Bednarek \& Bartosik~(2005) is low, i.e. between $\sim 0.8\%$ with the low level of soft background inside the nebula up to $\sim 3.2\%$ with the high level of soft background (additional infrared component). Therefore, in this case the upper limits reported in this paper cannot constrain the model yet.

\section{Conclusions}\label{sec:conc}

Our search for VHE $\gamma$-ray emission from the canonical pulsars PSR J0205+6449, PSR J2229+6114 and PSR J0218+4232, the only millisecond pulsar detected at $\gamma$-rays by the EGRET Telescope, was negative. The MAGIC observation of PSR J0205+6449 and its surrounding pulsar wind nebulae 3C 58 has not revealed any emission for E$>$110 GeV coming from the pulsar magnetosphere or the interaction between the pulsar wind and the surrounding medium. Possible TeV emission, as predicted by less conservative models for the pulsed $\gamma$-ray emission (see e.g.~\cite{Hirotani2007}), seems to be below the MAGIC sensitivity in the explored energy range. Also, MAGIC did not detect any $\gamma$-ray signal from the millisecond pulsar PSR J0218+4232 in the explored energy range above 70 GeV.  
These results lead us to conclude that both more sensitive and lower energy threshold Cherenkov telescopes are needed for the observation of these pulsars. 
The improved sensitivity of the MAGIC stereo-system with the new trigger system, should allow us to study these sources with a threshold $\sim$ 25-40 GeV and will be thus crucial for the study of the Northern sky pulsars. Similarly, the next year to be commissioned HESS II IACT should be able to reach an equivalent threshold for observations of pulsars in the Southern Sky. Also, observations in the MeV-GeV domain by Fermi~\cite{FermiCataloga}, possibly just overlapping around 40 GeV with ground-based observations should help to widen the understanding of high-energy $\gamma$-ray emission from the pulsar magnetosphere.

\section*{Acknowledgements}

We thank the IAC for the excellent working conditions at the ORM.
The support of the german BMBF, MPG and the YIP of the Helmholtz gemeinschaft, the Italian INFN, the Spanish MICINN, the ETH Research Grant TH 34/04 3 and the Polish MNiI Grant 1P03D01028 is gratefully acknowledged.


\end{document}